\documentclass{article}

\usepackage{graphicx}
\graphicspath{{figures/}}
\usepackage[T1]{fontenc}
\usepackage{bbm}
\usepackage{ccaption}
\usepackage{tkz-fct}
\usepackage{tikz,pgfplots,tkz-tab}
\usepackage{enumerate}
\usepackage{stmaryrd} 
\usepackage[toc,page]{appendix} 
\usepackage{amsmath,amssymb} 
\usepackage{url}
\usepackage{comment}

\newcommand{\B}{\operatorname{B}}
\newcommand{\Bin}{\operatorname{B}}
\newcommand{\Bbar}{\operatorname{C}}
\renewcommand{\ll}{\llbracket}
\newcommand{\rr}{\rrbracket}
\newcommand{\abs}[1]{\left\lvert#1\right\rvert}

\renewcommand{\le}{\leqslant}

\renewcommand{\ge}{\geqslant}

\newcommand{\Linf}{L_{\infty}}
\newcommand{\G}{\mathcal{G}}
\newcommand{\ind}{\ensuremath{\mathbbm{1}}}
\renewcommand{\P}{\ensuremath{\mathbb{P}}}
\newcommand{\E}{\ensuremath{\mathbb{E}}}

\newcommand{\N}{\ensuremath{\mathbb{N}}}
\newcommand{\R}{\ensuremath{\mathbb{R}}}

\newcommand{\C}{\ensuremath{\mathbb{C}}}

\newcommand\erf{\operatorname{erf}}
\newcommand\ud{\,\mathrm{d}}
\newcommand\Normal{\mathcal{N}}

\title{Effects of initial telomere length distribution on senescence onset and heterogeneity}

\date{\today}

\begin{document}

\maketitle

\centerline{Sarah Eug\`ene$^1$, Thibault Bourgeron$^2$, Zhou Xu$^{3}$}
{ \footnotesize $^1$Sorbonne Universit\'es, UPMC Universit\'e Pierre et Marie Curie, UMR 7598, Laboratoire Jacques-Louis Lions, F-75005, Paris, France and INRIA Paris, 2 rue Simone Iff, F-75012 Paris, France, email: Sarah.Eugene@inria.fr 

       $^2$\'Ecole Normale Sup\'erieure de Lyon, UMR 5569, Unit\'e de Math\'ematiques Pures et Appliqu\'ees, 69007 Lyon, France and INRIA Numed, 46 all\'ee d'Italie, 69007 Lyon, France, email: thibault.bourgeron@ens-lyon.fr, \emph{corresponding author}

       $^3$Sorbonne Universités, UPMC Univ Paris 06, CNRS, UMR 8226, Laboratoire de Biologie Moléculaire et Cellulaire des Eucaryotes, Institut de Biologie Physico-Chimique, 75005 Paris, France, email: zhou.xu@ibcp.fr, \emph{corresponding author} }

\vspace*{\fill}
       
\textbf{Abstract.} Replicative senescence, induced by telomere shortening, exhibits considerable asynchrony and heterogeneity, the origins of which remain unclear. Here, we formally study how telomere shortening mechanisms impact on senescence kinetics and define two regimes of senescence, depending on the initial telomere length variance. We provide analytical solutions to the model, highlighting a non-linear relationship between senescence onset and initial telomere length distribution. This study reveals the complexity of the collective behavior of telomeres as they shorten, leading to senescence heterogeneity. 

\vspace*{\fill}

\medskip


\medskip







\newpage

\section{Introduction}

\begin{figure}[htbp]
\centering
\includegraphics[scale=1]{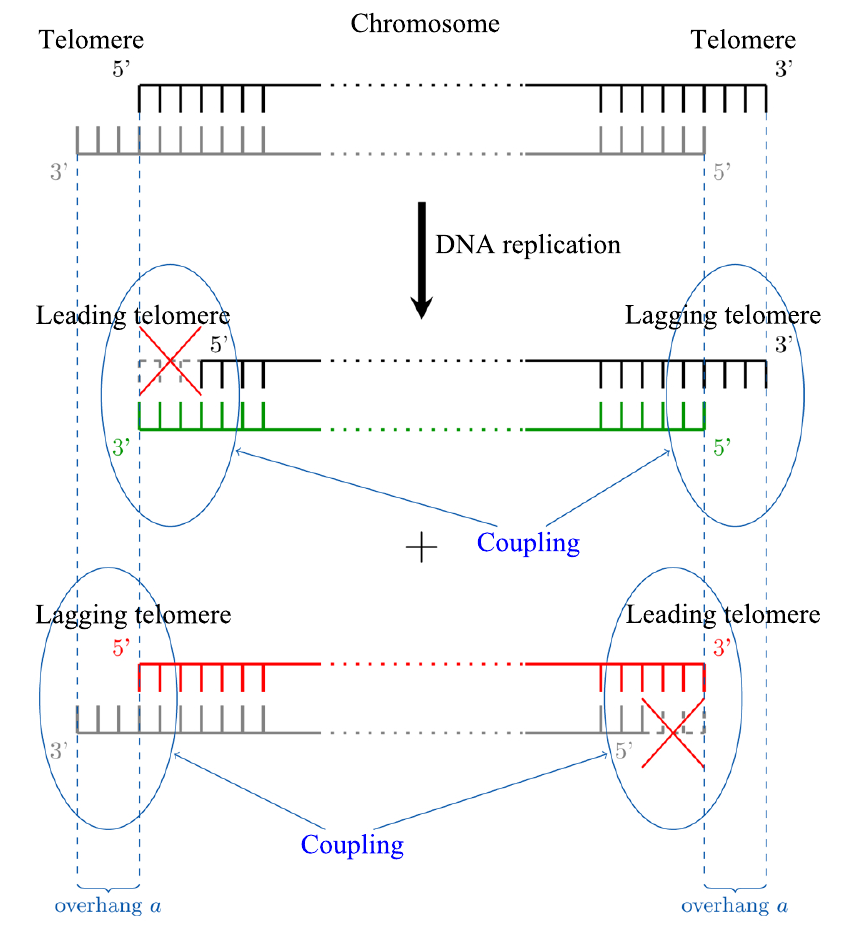}
\caption{\footnotesize{Scheme of a chromosome bearing two telomeres and undergoing replication. Telomeres end with a $3$' overhang of length $a$ (measured to be $5-10$ nucleotides in yeast \cite{Soudet}, chosen here as $a=1$ or $7$ for theoretical or numerical purposes, respectively). After DNA replication, each telomere generates, through either the leading or the lagging strand replication machineries, two new telomeres of different lengths. The coupling effect between the two ends of the same chromosome imposes that only one of the two is shortened while the other retains the parental length.}}
\label{ffig1}
\end{figure}

Telomeres, the ends of eukaryote chromosomes, are poised in a dynamic equilibrium controlled by two processes: limited telomere shortening at each cell division and elongation by telomerase, a dedicated holoenzyme able to generate \emph{de novo} telomere sequence.  When telomerase is not expressed, as in human somatic cells, or is experimentally mutated in model organisms such as \emph{Saccharomyces cerevisiae} \cite{Lundblad}, telomeres only shorten and after many divisions the cell enters replicative senescence, a permanent cell cycle arrest induced by short telomeres that elicit a DNA damage response. Replicative senescence is implicated in organismal ageing and is a potent barrier to cancer emergence, but its remarkable asynchrony and heterogeneity remain a challenge for investigating the exact relationship between initial telomere length distribution and senescence onset. \\
Telomere shortening is the unavoidable consequence of the end-replication problem \cite{Olovnikov, Watson, Soudet}. In most examined species, telomeres end with a $5$' to $3$' singled-stranded DNA overhang (Fig.~\ref{ffig1}) \cite{HG99, HB89, Klobutcher, Makarov, EW97, Raices, Riha, Wellinger}. When the replication fork reaches the end of the chromosome, the processing of the last Okazaki fragment leaves a gap at the lagging strand, which recreates the single-stranded overhang of the parental telomere (Fig.~\ref{ffig1}). On the leading strand, after replication, complex maturation steps involving resection and fill-in also regenerate the overhang structure \cite{Larrivee, Faure, Chai, WTL12, Soudet}. Regardless of these maturation steps, the leading strand template for replication is shorter than the lagging strand one, thus generating after replication two new telomeres of different lengths, one unchanged compared to the parental telomere and the other shorter by exactly the length of the overhang, as illustrated in Fig.~\ref{ffig1}. Previous mathematical models of telomere shortening also based on the end-replication problem~\cite{levy,arino,olofsson1999, arkus} did not consider the maturation of the leading strand telomere that generates a 3'-end overhang identical to the one on the lagging strand. This maturation step is widely conserved throughout species with the notable exception of angiosperm plants that display a blunt end at the leading telomere~\cite{Riha}. We also note that other mathematical models examined higher level structures such as t-loops~\cite{griffith1999, rodriguez}, or additional telomere states or breaking mechanisms~\cite{kowald, rubelj,proctor2002, proctor2003}, such as damage due to oxidative stress~\cite{von2002}. In \emph{S. cerevisiae,} however, oxidative stress does not significantly alter telomere length~\cite{romano2013} and the end-replication problem is the main mechanism of telomere shortening.\\
Consistently, on average, telomeres shorten at a constant rate of exactly half of the overhang length per division. However, while studying the average is informative of the global regulation and homeostasis of telomere length, it misses important contributions of the asymmetry of telomere replication mechanism to the overall telomere length distribution and to the heterogeneity of the onset of senescence. Taking this asymmetry into account, the shortening of a telomere in a cell lineage, defined as a random succession of mitotically related cells \cite{Xu15}, is probabilistic and follows a Bernoulli process. Additionally, if the two ends of a given chromosome are considered together, the $3$'-end at one telomere belongs to the same DNA strand as the $5$'-end on the other telomere of the same chromosome, implying that the asymmetry at one telomere is inverted compared to the other (Fig.~\ref{ffig1}). We define this relationship between the two ends of the same chromosome as a coupling mechanism, which adds another layer of constraint and  will also be modeled here. \\
In this article, we study the consequences of the asymmetry and the coupling on the distribution and the dynamics of telomere length in two distinct phases: at steady state in the presence of telomerase and in a strictly shortening phase without telomerase. We show that the robustness of telomerase recruitment impacts on the variance of the steady-state distribution of telomere length. In turn, this variance defines different regimes of senescence. In a regime of low initial variance, senescence onset cannot be linearly inferred from the average telomere length or even the length of the shortest telomere and we provide an asymptotic expansion to account for this phenomenon. In contrast, a high variance implies a linear correlation between the initial shortest telomere and senescence onset. We provide analytical solutions to the different models we describe and suggest applications for the inference of the initial telomere length distribution from experimental measurements of senescence onset.

\section{Telomeres evolving with telomerase}

\begin{figure}[htbp]
\centering
\includegraphics[scale=0.4]{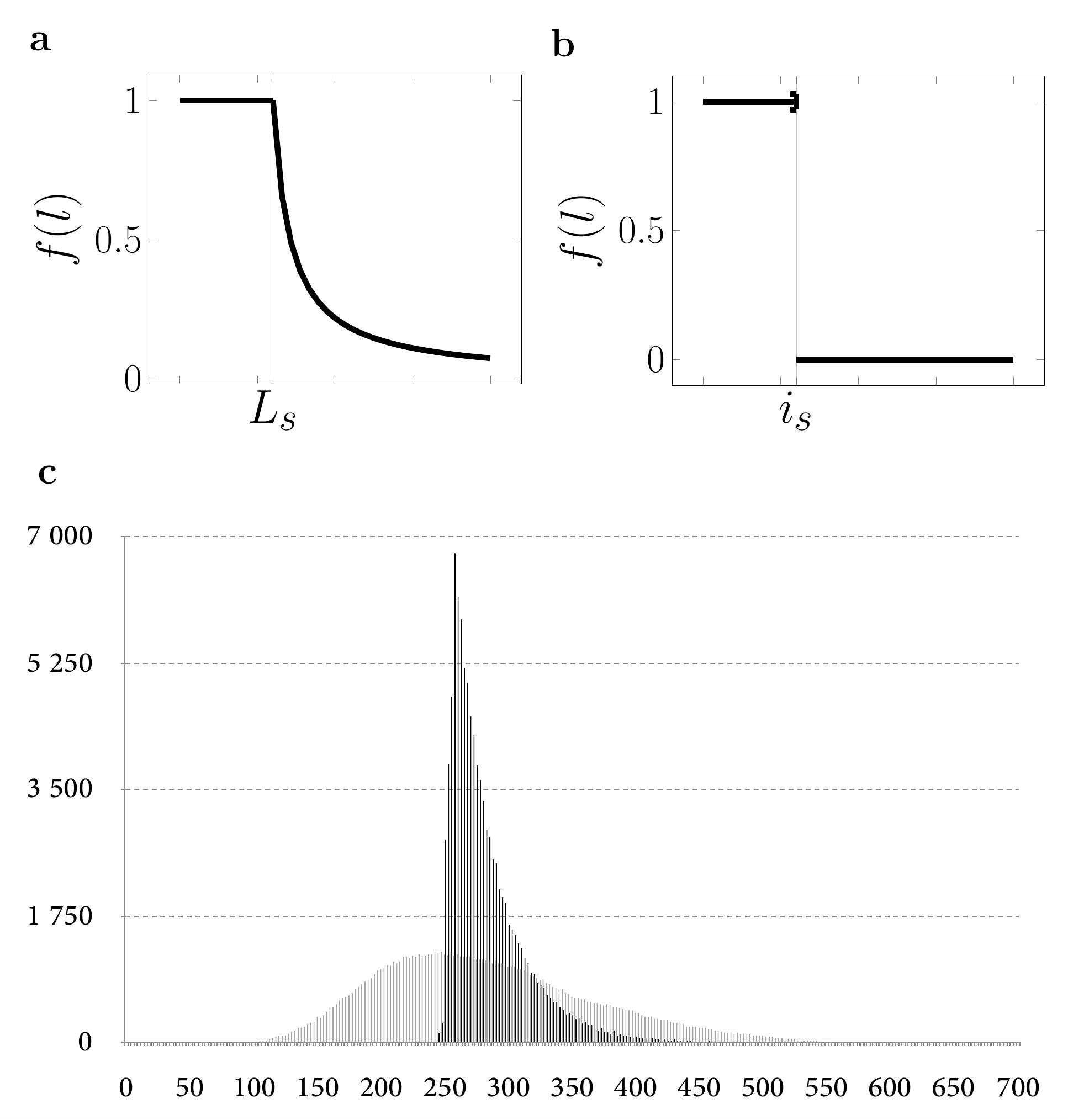}
\caption{\footnotesize{Steady-state telomere length distribution in the presence of telomerase. (a) and (b) Probability of recruitment and action of telomerase, as modeled from \cite{T04} with a length threshold $L_s$ or simplified with a sharp switch occurring at $i_s$. (c) Simulation of the steady-state distribution of telomere length using either (a) (black) or (b) (grey) to describe telomerase recruitment. $i_s$ was set so as to reach the same mean in the steady-state distribution (Appendix \ref{app:choiceis}).}}
\label{ffig2}
\end{figure}

We first describe the most general model, corresponding to a lineage of haploid yeast cells dividing in the presence of active telomerase. The two telomeres of a given chromosome at generation $n$, called $(L^1_n, L^2_n)$, are coupled as defined above and the $32$ telomeres of the cell shorten according to a Bernoulli random variable $\B_n$ of parameter $1/2$: if $\B_n=1$, then $L^1_n$ is shortened by $a$ nucleotides, whereas $L^2_n$ is preserved, and conversely if $\B_n=0$. Telomerase adds new telomere sequences preferentially to shorter telomeres \cite{T04, BC09}, behavior that we capture by introducing $\Bbar_n^i$, $i\in \{1,2\}$, Bernoulli random variables of parameter $f(L_n^i)$, according to~\cite{Xu13}, where $f$ has the shape shown in Fig.~\ref{ffig2} (a) and $L^i_n$ is the length of the telomere at the extremity $i$ before replication. The shape of $f$ is such that below a length threshold $L_s$, the Bernoulli random variable $\Bbar_n^i$ equals $1$, that is telomerase is always active. For a telomere longer than $L_s$, the probability of $\{\Bbar_n^i = 1\}$ decreases to zero, meaning that the longer the telomere, the less likely it is to be elongated by telomerase. Since the number of nucleotides added by telomerase is independent of the length of the telomere \cite{T04, Xu13}, we introduce $\G^1_n$ and $\G^2_n$ two independent geometric random variables of parameter $p$, independent of all the other quantities (including $L_n^1, L_n^2$),  which correspond to the number of nucleotides added by telomerase. As a result, for any given chromosome the $\mathbb{N}^2$-valued process $(L^1_n, L^2_n)$ follows:

\begin{equation}\label{eq:coupledModel}
\begin{pmatrix}
L_{n+1}^1\\
L_{n+1}^2
\end{pmatrix}
=
\begin{pmatrix}
(L_{n}^1 - a\cdot \B_n)^+ + \Bbar^1_n \cdot \G^1_n\\
(L_{n}^2 - a\cdot (1-\B_n))^+ + \Bbar^2_n \cdot \G^2_n
\end{pmatrix},
\end{equation}
where $x^+=\max(0,x)$ is $x$ if $x>0$ and $0$ otherwise. Telomere length is nearly always positive because, in the presence of telomerase, a telomere shorter than $L_s>0$ is elongated with probability $1$, while in the absence of telomerase, senescence is triggered before the short telomere reaches $0$ \cite{Bourgeron}.
Using the same Bernoulli random variable $\B_n$ for $L_n^1$ and $L_n^2$ mathematically defines the coupling between the two telomeres. \\
To characterize the steady state of telomere length distribution, we focus on one telomere---because as an approximation, telomeres of different chromosomes are assumed to be independent \cite{SB88}---, and consider the projection of the first coordinate of a chromosome in order to compute its equilibrium. We will analyze the coupling effect in more depth in the second regime without telomerase. Our model thus becomes:
\begin{equation}\label{eq:modelcomplet}
L_{n+1} = (L_{n} - a\cdot \B_n)^+ + \Bbar_n \cdot \G_n
\end{equation}
where $L_n$ is the length of a given telomere, $L_{n+1}$ the length of one of the two daughter telomeres, $\G_n$ a geometric random variable of parameter $p\in (0,1)$. An averaged version of this model has been studied in~\cite{Xu13,DD13} and used in \cite{Bourgeron}, where instead of being stochastic, telomere shortening was chosen to be deterministic with a constant value of $a/2$. 
To make our computations fully explicit without betraying the principles of the biological mechanism, instead of $f$, we consider a sharp threshold at a value $i_s$ (Fig.~\ref{ffig2}b). Our model becomes: 
\begin{equation} \label{eq:model}
L_{n+1} = (L_n - a \cdot \B_n)^{+} + \G_n \cdot  \ind_{\{L_n \le i_s\}}
\end{equation}

Independently of the value of $i_s$, the Markov chain $(L_n)_{n\ge0}$ defined by \eqref{eq:model} has a unique equilibrium distribution $\Linf$ and the generating function of $\Linf$ is characterized by the equality:
\begin{equation} \label{eq-gene}
\begin{split}
\forall u \in \C,\abs{u} \le 1, \qquad (u^a-1) &(1-uq) \; \E\left(u^{\Linf} \right) \\
& = - q (1-u) (u^a+1) S_{i_s}(u) + p S_{a-1}(1) u^a - p S_{a-1}(u),
\end{split}
\end{equation}
where $\pi_j = \P(\Linf=j)$, $S_k(u) = \E\left(u^{\Linf} \; \ind_{ \{\Linf \le k \}} \right) = \sum_{j=0}^{k} \pi_j \; u^j$ and $q=1-p$.

Appendix~\ref{app:initialstate} gives a proof of this result and explains how to get a fully explicit expression for the distribution $\Linf$ from equation \eqref{eq-gene}. This calculation reveals how the parameters $a$ and $p$ affect the steady-state distribution.

The choice of the value of $i_s$ according to biological experiments is explained in~Appendix~\ref{app:choiceis}. We find that the variance of the steady-state telomere length distribution obtained using the simplified model \eqref{eq:model} is significantly smaller than the one with the complete model \eqref{eq:modelcomplet} (Fig. \ref{ffig2}c in black and grey, respectively; $37$ bp as compared to $101$ bp), demonstrating that the residual recruitment of telomerase to rather long telomeres strongly contributes to the spread of the steady-state distribution of telomere length. In turn, the variance of this distribution is critical for determining the onset of senescence and its heterogeneity, as we show below. Thus, the mode of recruitment and activation of telomerase, dependent on the biochemical properties of the holoenzyme and on its interactions with telomeric proteins 
~\cite{WellingerZakian}, controls key features of senescence once telomerase is removed.


\section{Telomeres evolving without telomerase}

We then analyze the consequences of the steady-state distribution on the onset of senescence, meaning the number of generations undergone by a given cell lineage until it enters senescence. We simply call it time of senescence, denoted by $T$. One practical goal of this section is to derive the parameters of the initial distribution from the time of senescence, which is useful for experimentalists. In senescing cells, telomerase is inactive and when the shortest telomere reaches a threshold $S$, the cell enters replicative senescence and stops dividing \cite{Abdallah, Lundblad, Hemann, Zou, Armanios}. A haploid yeast cell has $16$ chromosomes and thus $32$ telomeres. Mathematically, we consider the vector $(L^1_n, L^2_n,\dots, L^{32}_n)$ of these $32$ telomere lengths at generation $n$. Because each chromosome behaves independently~\cite{SB88}, we can start by studying one chromosome and the behavior of the $16$ will easily follow. More precisely, the vector $(L^1_n, L^2_n,\dots, L^{32}_n)$ can be seen as a family $(X_n^i,Y_n^i)_{1\le i\le16}$ of $16$ independent identically distributed couples each representing the two telomeres of a chromosome, with $(X_0^i,Y_0^i)\overset{dist.}{\sim}\Pi$. The time of senescence is mathematically expressed as:
\[T = \inf \left\{ n \ge 0, \min_{1\le i \le 16} \left[ \min(X_n^i,Y_n^i)\right] < S \right\}.\]

\noindent\textbf{Normalizations.} 
As telomeres can only shorten, we consider the shortening length to be $a=1$. For numerical estimations of the time of senescence, we will divide our results by $a=7$ to obtain biologically relevant values. Moreover, we can choose $S=0$ by simply translating the initial state by $S$. These assumptions are made in all following calculations unless stated otherwise.\\

\noindent\textbf{Distribution of the Time of Senescence.}
Under these normalization conditions, we find that the distribution of the random variable $T$ is fully explicit:
\begin{equation} \label{eq:PTgen}
\P(T> n) = \left[ \sum_{k+l\ge n}\Pi(X_0 = k,Y_0=l) \;2^{-n}\sum_{t= n - l }^{k} \binom{n}{t} \right] ^{16}.
\end{equation}

In particular, its expectation can be written as a function of $\pi$ as follows:
\begin{equation} \label{eq:expT}
\E(T) =\sum_{n = 0 }^{\infty} \left[ \sum_{k+l\ge n}\Pi(X_0 = k,Y_0=l)\;2^{-n}\sum_{t= n - l }^{k} \binom{n}{t} \right] ^{16}.
\end{equation}

See Appendix \ref{app:expected} for the proof. Because of the difficulty to invert this formula, we choose to study separately the influence of the mean and the variance of the initial state on the time of senescence. Thus, we first consider a deterministic and constant initial state $X_0=x_0$:
\[ k=1,\dots,32, \quad L_0^k = \E(\Linf) =: x_0\]

We define $T^1_{x_0}$ as the first time one of two coupled telomeres reaches zero both starting from $x_0$, and $T_{x_0}$ as the time of senescence of the whole cell when the initial state is constant and equals $x_0$.

Almost surely, $x_0 \le T^1_{x_0} \le 2 \, x_0$, this implies that $\P(T^1_{x_0}> n)  = 0$ for $n\ge2 x_0$, and $\P(T^1_{x_0}\ge n) = 1$ for $n < x_0$.
For $x_0 \le n \le 2x_0-1$, the law of $T^1_{x_0}$ is given by:
\[\P(T^1_{x_0}> n) = 2^{-n} \sum_{k = n-x_0}^{x_0} \binom{n}{k}.\]

The expected time of senescence is then:
\begin{equation} \label{eq:expconstant}
\E(T_{x_0}) = x_0 + \sum_{n = x_0 }^{2 x_0-1}\left[ \P(T^1_{x_0}> n) \right] ^{16}
= x_0 + \sum_{n = x_0 }^{2 x_0-1}\left[ 2^{-n} \sum_{k = n-x_0}^{x_0} \binom{n}{k}\right] ^{16}.
\end{equation}

We then perform an asymptotic expansion of $\E(T_{x_0})$ for large values of $x_0$, which is numerically justified (Appendix~\ref{app:choiceis} and Fig.~\ref{ffig2}c). At the first order the mean behavior prevails:
\begin{equation}
\E(T^1_{x_0}) \sim 2 x_0.
\end{equation}

Concerning the second order, we obtain the following convergence in distribution:

\begin{equation} \label{eq:loideT1}
\frac{2 x_0 - T^1_{x_0}}{\sqrt{x_0}} \mathrel{\mathop{\longrightarrow}^{dist.}_{x_0 \to +\infty}} \abs{N}, \qquad \text{with } N \overset{dist.}{\sim} \Normal(0,2).
\end{equation}
See Appendix \ref{app:asymp} for proofs of these results.

\medskip

The asymptotic development of the time of senescence for one chromosome $T_{x_0}^1$ allows us to derive an approximation of the expected time of senescence \eqref{eq:expconstant} by replacing the law of $T_{x_0}^1$ by its asymptotic~\eqref{eq:loideT1}:
\begin{equation} \label{eq:expasymp}
\E(T_{x_0}) \approx x_0 + \sum_{n = x_0 }^{2 x_0-1} \left[ \P(2 x_0 -  \sqrt{x_0}  \abs{N} > n)\right] ^{16} = x_0 +\sum_{k=0}^{x_0-1}  \left[  \erf \left( \frac{k}{2 \sqrt{x_0}}\right) \right]^{16},  
\end{equation}
where $\erf$ is the error function defined as:
\[
{\displaystyle \erf x={\frac {2}{\sqrt {\pi }}} \int _{0}^{x}e^{-t^{2}} \ud t.}
\]

\begin{figure}[htbp]
\centering
\includegraphics[scale=0.45]{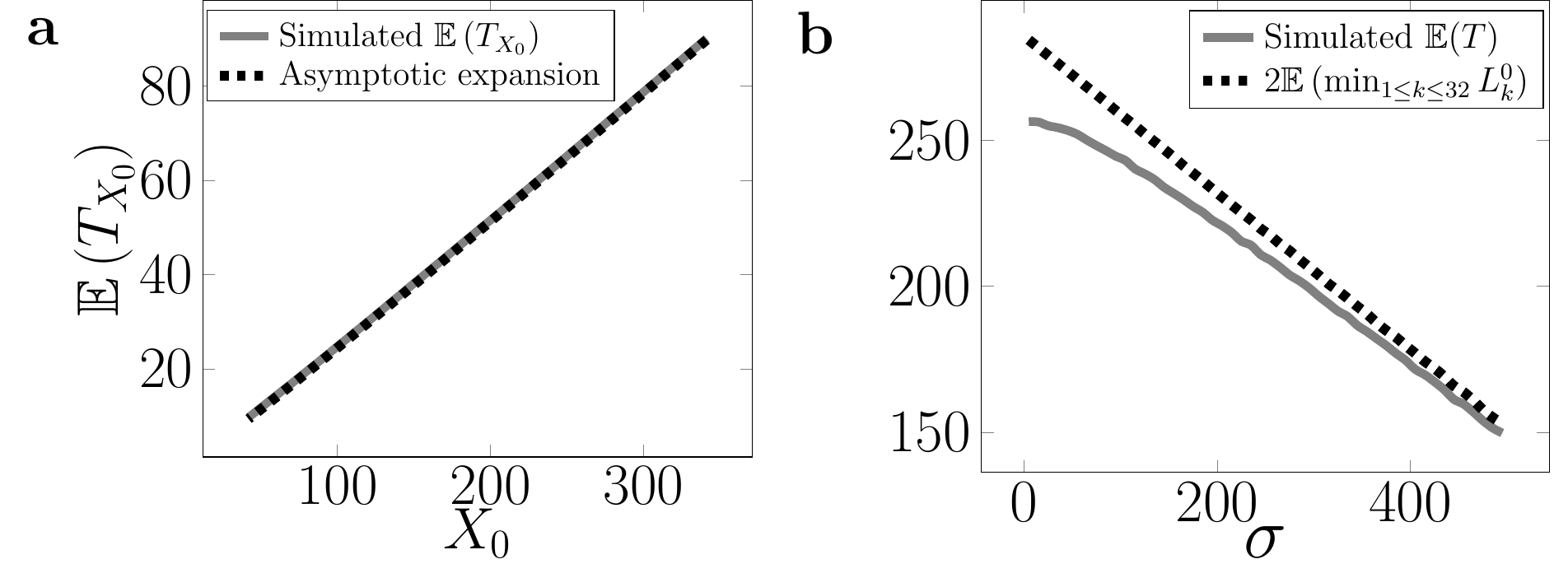}
\caption{\footnotesize{Distinct effects of the mean and variance of the initial distribution on theoretical expressions and numerical simulations of the time of senescence. (a) Starting from a constant distribution $\E(\Linf):=x_0$, the asymptotic expansion in equation \eqref{eq:expasymp} is computed and compared to numerical simulations ($1000$ independent simulations). (b) Starting from a uniform distribution of variance $\sigma$ and mean $\E(\Linf)$, the time of senescence is computed using equation~\eqref{eq:minimpact}, which takes only the mean behavior of the initial shortest telomere into account, and compared to numerical simulations ($1000$ independent simulations).}}
\label{ffig3}
\end{figure}

We find that the expansion~\eqref{eq:expasymp} is hardly distinguishable from the theoretical process~\eqref{eq:expT}, as shown in simulations (compare grey and dashed black lines in Fig.~\ref{ffig3} (a)) and can thus be directly used to estimate the mean of the initial state in experimental studies.

\noindent\textbf{Influence of the initial variance on the time of senescence.} 
Now, to study only the influence of the initial variance, we consider that each initial telomere is uniformly distributed in the interval $[\E(\Linf) - \sigma, \E(\Linf) + \sigma]$ and simulate the expected time of senescence as a function of $\sigma$ (Fig.~\ref{ffig3}b). 
When $\sigma$ has large values, there is a higher probability that the initial shortest telomere of ($L^1_0,\dots,L^{32}_0)$ is far from the mean $\E(\Linf)$ and, thus, that it remains the shortest one until senescence. We therefore expect, for large enough value of $\sigma$ (Fig. 3b), that the time of senescence is asymptotically equivalent  to the time when the initial shortest telomere reaches zero. As, on average, the number of steps for a simple random walk starting from $M= \E \left[\min_{1\le k\le 32} L_k^0\right]$ to reach zero is $M/(1/2)=2M$, we expect the following result:
\begin{equation} \label{eq:minimpact}
\lim_{\sigma\to \E(\Linf) }  \E(T)= 2 \; \E \left[\min_{1\le k\le 32} L_k^0\right].
\end{equation}
We indeed find this asymptotic behavior by simulations (Fig.~\ref{ffig3}b), highlighting two regimes that depend on the initial variance. If $\sigma$ has a small value, which is close to the deterministic initial state studied above (equation~\eqref{eq:expasymp}), the time of senescence is much smaller than expected by just considering the shortest telomere because of the coupling effect. If $\sigma$ has a large value, the time of senescence is mainly determined by the shortening of the initial shortest telomere.

\begin{figure}[htbp]
\centering
\includegraphics[scale=0.4]{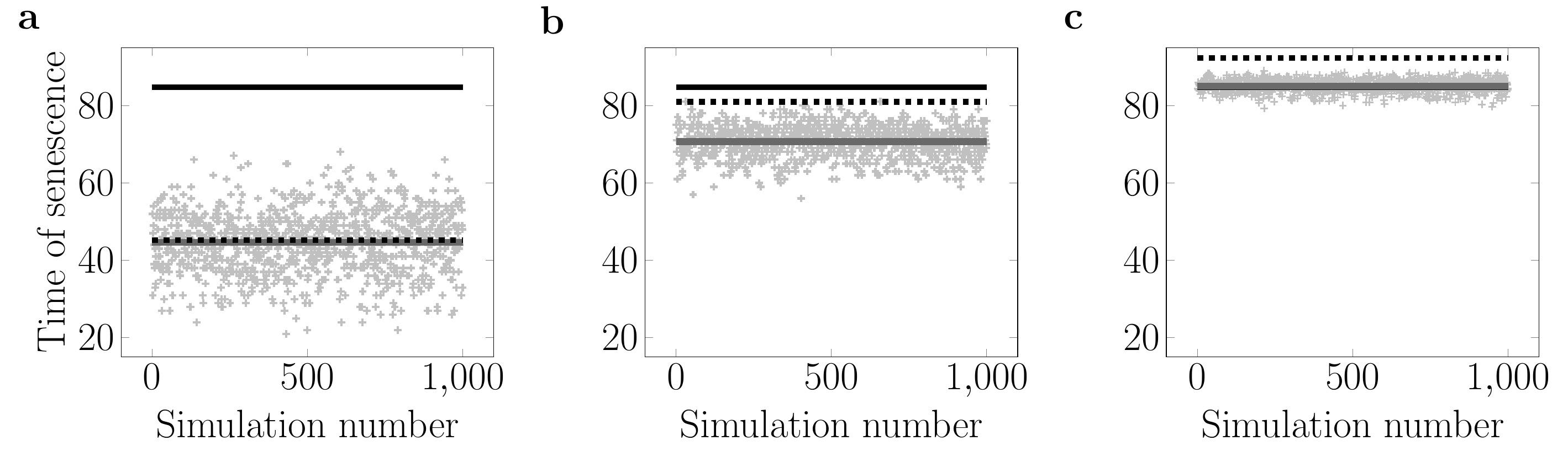}
\caption{\footnotesize{Comparison between simulated times of senescence (grey dots) and predictions from equations~\eqref{eq:minimpact} (dashed black lines) and~\eqref{eq:expasymp} (black lines). (a) 32 telomere lengths are randomly drawn from a biologically relevant distribution with a high variance (Fig.~\ref{ffig2}c, grey distribution) and the time of senescence is simulated to give one data point (grey dot). This process is repeated $1000$ times and compared to the two predictions. The grey line represents the average simulated time of senescence. (b) and (c): as in (a), but starting with an intermediate level of variance for the initial telomere length distribution or no variance at all, respectively.}}
\label{ffig4}
\end{figure}

We next ask which of these two regimes can be observed in simulated times of senescence from the telomere distributions described in Fig.~\ref{ffig2}, which are biologically more relevant than the previous distribution models. To do so, we simulate $1000$ individual lineages of senescing cells and record their time of senescence, starting by randomly drawing their 32 telomeres from the biologically relevant distribution (large variance, in grey in Fig.~\ref{ffig2}c), from the simplified distribution (intermediate variance, in black in Fig.~\ref{ffig2}c) or from a constant distribution (no variance) (Fig.~\ref{ffig4}). We then compare these simulated times of senescence with those predicted either from the mean behavior of the shortest telomere (equation~\eqref{eq:minimpact}, dashed black lines in Fig.~\ref{ffig4}) or the asymptotic expansion on the mean of the initial distribution (equation~\eqref{eq:expasymp}, black lines in Fig.~\ref{ffig4}). The biologically relevant distribution gives simulated times of senescence that are fully predicted by computing the mean behavior of the average initial shortest telomere (Fig. ~\ref{ffig4}a, compare grey and dashed black lines which are superimposed). In contrast, the constant distribution leads to a senescence onset dictated by the asymptotic expansion (Fig.~\ref{ffig4}c, compare grey and black lines), consistent with the results in Fig.~\ref{ffig3}a. The simplified distribution produces an intermediate result where the mean behavior of the shortest telomere and the asymptotic expansion lead to similar predictions, which lay close to the simulated times of senescence (Fig.~\ref{ffig4}b, compare grey, black and dashed black lines). These results show that defining two senescence regimes depending on the initial variance of telomere length distribution is critical for understanding the relevant dynamics of telomere shortening leading to senescence.

\section{Conclusion}

In summary, in this article, we isolated all the sources of fluctuations of the time of senescence that are dependent on telomere length. To do so, we modeled several molecular mechanisms that contribute at various levels to telomere length distribution and dynamics in \emph{S. cerevisiae}, where they are the most exhaustively and quantitatively described. Among these mechanisms, we found that the asymmetry of telomere replication and the coupling between the two telomeres belonging to the same chromosome significantly contribute to senescence heterogeneity and we formally established their links. We also showed that the mode and robustness of telomerase recruitment control the variance of the steady-state telomere length distribution, which in turn defines two senescence regimes. With a low initial variance, the time of senescence is non-linearly related to the initial mean telomere length. In contrast, a high initial variance leads to a major role of the initial shortest telomere in controlling senescence. Because natural telomere length distributions can vary considerabely, even within a species, we suggest that depending on the initial variance, the two regimes we describe may operate at the same time during senescence. As the core mechanisms modelled here are conserved in most eukaryotes, we expect that our conclusions should also, in principle, apply to telomere-dependent senescence in human cells, although additional factors and mechanisms also contribute to senescence heterogeneity \cite{griffith1999, rodriguez, proctor2002}. This work uncovers a new layer of complexity in the relationship between senescence onset and telomere shortening explained by the asymmetry and coupling mechanisms, and proposes methods for assessing the time of senescence or conversely inferring parameters of the initial telomere length distribution.
\newpage

%
%
%
%
%


\begin{appendices}
\setcounter{equation}{0}
\setcounter{figure}{0}
\setcounter{table}{0}
\renewcommand{\theequation}{\Alph{section}.\arabic{equation}}
\renewcommand{\thefigure}{\Alph{section}.\arabic{figure}}
\renewcommand{\thetable}{\Alph{section}.\arabic{table}}

\section{Steady state of telomeres evolving with telomerase} \label{app:initialstate}

First, we prove that both Markov chains defined by \eqref{eq:modelcomplet}, \eqref{eq:model} are ergodic and, second, we derive formula \eqref{eq-gene}. The ergodicity is a direct consequence of Foster-Lyapunov criteria (Corollary 8.7 p. 214 in~\cite{Philippe} or Proposition 1.3 in~\cite{hairer}, for instance). The chains are time-homogeneous and, for $L_0 > a$:
\[ \E(L_1 - L_0) = \E(-a \cdot \B_0 + \Bbar_0\cdot \G_0) = -a/2 + f(L_0) \; \E(\G_0), \]
and this last quantity is negative for large enough values of $L_0$ because $f$ tends to zero, either $f$ is $f(l)= (1+\beta (l-L_s))^{-1}$ or $f(l)=\ind_{l\le i_s}$. This proof of ergodicity works for any function $f$ having a limit $l$ at infinity which satisfies $l < \frac{a}{2 \E(\G_0)}$.

Moreover, the Markov chains $(L_n)$ defined by \eqref{eq:modelcomplet}, \eqref{eq:model} are irreducible and aperiodic. Therefore, in both cases, there exists a unique equilibrium distribution, denoted $\Linf$. For the model~\eqref{eq:model}, it is characterized by the fact that $\Linf$ and $(\Linf - a\cdot \B_0)^{+} + \G_0  \ind_{\{\Linf \le i_s\}}$ have the same probability generating function.



To establish formula \eqref{eq-gene} we distinguish the three regimes: $\Linf < a\cdot \B_0$, $a\cdot \B_0 \le \Linf \le i_s$, $\Linf >i_s$, and the cases $\B_0=0$ or $\B_0=1$.
For $u$ such that $\abs{u}\le 1$, we obtain:
\begin{equation} \label{eq-calcLinf}
\begin{split}
2  \E\left(u^{\Linf} \right)
&= 2 \E\left(u^{(\Linf-a\cdot \B_0)^+ + {\G_0} \cdot \ind_{\Linf \le i_s}} \right) \\
&= 2 \E \left(u^{\G_0} \ind_{\{ \Linf < a\cdot \B_0 \}}\right) + 2 \E \left(u^{\Linf -a\cdot \B_0 + \G_0} \ind_{\{ a\cdot \B_0 \le \Linf \le i_s \}}\right) + 2 \E \left(u^{\Linf -a\cdot \B_0} \ind_{\{ \Linf > i_s \}}\right)  \\
&= \E \left(u^{\G_0}\right) \bigg[ \P (\Linf < a) + \P(\Linf<0) \bigg]
+ \E \left(u^{\G_0}\right)
\bigg[ \E\left( u^{\Linf -a} \ind _{\{a\le\Linf \le i_s \}}\right) \\
&+ \E \left( u^{\Linf} \ind _{\{0\le\Linf \le i_s \}}\right) \bigg]
+ \bigg[ \E\left( u^{\Linf -a} \ind _{\{\Linf >i_s \}}\right)
+ \E \left( u^{\Linf} \ind _{\{\Linf > i_s \}}\right) \bigg] \\
&= \E \left(u^{\G_0}\right) S_{a-1}(1)
+  \E \left(u^{\G_0}\right) 
\bigg[ u^{-a} \left(S_{i_s}(u)-S_{a-1}(u) \right)
+ S_{i_s}(u) \bigg] \\
&+  
\bigg[ u^{-a} \left( \E\left(u^{\Linf}\right) - S_{i_s}(u) \right)
+ \left( \E\left(u^{\Linf}\right) - S_{i_s}(u) \right) \bigg] .
\end{split}
\end{equation}

From \eqref{eq-calcLinf}, we get:
\[
\E\left(u^{\Linf} \right) \left(1-u^{-a} \right)
= S_{i_s}(u) \left( \E\left(u^{\G_0} \right) -1 \right) \left(1+u^{-a} \right)
+ \E\left(u^{\G_0} \right) \left( S_{a-1}(1) - u^{-a} S_{a-1}(u) \right).
\]

As the probability generating function of a geometric distribution is explicitly given by $\E\left(u^{\G_0}\right) = p (1-uq)^{-1}$, with $q=1-p$, we obtain \eqref{eq-gene} after multiplication by $(1-uq) u^a$. 

In order to compute the $\pi_k$ for all $k$, we identify the coefficients of the power series of each side of \eqref{eq-gene} distinguishing cases for the values of $k$.  For simplicity we set $\pi_k = 0$ for $k<0$. Using the identities: $(u^a-1)(1-uq) = -q u^{a+1} + u^a + qu -1$, $(u-1)(u^a+1) = u^{a+1}-u^a + u - 1$, we get:
\begin{equation} \label{eq-gene-2}
\begin{split}
\sum_{k=0}^\infty &  \left( -q \pi_{k-a-1} + \pi_{k-a} + q \pi_{k-1} - \pi_k \right) u^k 
= \sum_{k=0}^\infty \pi_k  \left( -q u^{k+a+1} + u^{k+a} + q u^{k+1} -  u^k \right) \\
&= (-q u^{a+1} + u^a + qu -1)  \E\left(u^{\Linf} \right) \\
&= q (u^{a+1}-u^a + u - 1) S_{i_s}(u) + p S_{a-1}(1) u^a - p S_{a-1}(u) \\
&= q \bigg[ \sum_{k=a+1}^{i_s+a+1} \pi_{k-a-1} \; u^k - \sum_{k=a}^{i_s+a} \pi_{k-a} \; u^k  + \sum_{k=1}^{i_s+1} \pi_{k-1} \; u^k  - \sum_{k=0}^{i_s} \pi_{k} \; u^k \bigg] \\
& + p S_{a-1}(1) \; u^a - p \; \sum_{k=0}^{a-1} \pi_{k} \; u^k.
\end{split}
\end{equation}

The following table gives the recurrence relations obtained after identification in the coefficients of \eqref{eq-gene-2}.
\[\begin{array}{c|l}
k & \text{relation} \\ \hline
\ll 0,a-1\rr       &  \emptyset \\
a                 & \pi_a = 2 \frac{1-p}{p}\pi_0 - \sum_{k=1}^{a-1} \pi_k \\
\ll a+1,i_s\rr     & \pi_k =  -2 \frac{1-p}{p} \pi_{k-a-1} + \frac{2-p}{p} \pi_{k-a} \\
i_s + 1           &  \pi_{i_s+1} = - 2q \pi_{i_s-a} + (1+q) \pi_{i_s+1-a} \\
\ll i_s+2,i_s+a\rr & \pi_{k} = - 2q \pi_{k-a-1} + (1+q) \pi_{k-a} + q \pi_{k-1} \\
i_s + a + 1       & \pi_{i_s+a+1} = - 2q \pi_{i_s} + \pi_{i_s+1} + q \pi_{i_s+a} \\
> i_s + a + 1     & \pi_{k} = - q \pi_{k-a-1} + \pi_{k-a} + q \pi_{k-1}
\end{array}
\]

Note that for $k$ such that $0\le k \le a-1$, the identification gives no information on the values of the $\pi_k$. These formulas show that all the $\pi_k$ depend linearly on the $a$ first states, $\pi_k$, $k=0,\dots,a-1$. Now, we indicate how to compute these $a$ first values. 


\medskip

\medskip


Hence, using~\eqref{eq-gene}, the generating function of $\Linf$ is only a function of the $a$ first states. Dividing~\eqref{eq-gene} by $(1-u)(1-uq)$, we can find $\psi: \R^a \times [0,1] \to \R$ linear in the first $a$ coordinates such that:
\begin{equation}
\label{eq:psi}
(1+ \dots + u^{a-1}) \,\E(u^{\Linf})= \psi(\pi_0, \dots, \pi_{a-1},u).
\end{equation}

The $a-1$ roots of $R(u) = 1+\dots +u^{a-1}$ are the $u_k = e^{2i\pi k/a}$ for $1\le k \le a-1$, which are in the unit disk. Therefore, the vector $(\pi_0,\dots,\pi_{a-1})$ is solution of the system:
\begin{equation}
\label{eq:system}
\psi(\pi_0,\dots, \pi_{a-1},u_k) = 0 \text{ for } 1\le k \le a-1
\end{equation}
where $\pi_0$ is, as usual determined, by the normalization condition. If the $a\times a$system~\eqref{eq:system} to which we added the normalization condition is invertible, then there exists a unique solution $(\pi_0,\dots,\pi_{a-1})$. Having the vector $(\pi_0,\dots,\pi_{a-1})$, the generating function $\E(\Linf)$ follows from~\eqref{eq:psi}. 

Finally, we want to explicitly determine the $\pi_k$ for $k\ge a$ .The previous table gives homogeneous linear recurrence relations with constant coefficients. For instance $\pi_k$, for $k$ in $\ll a+1, i_s \rr$, is a linear combination (independant of $k$) of the $k$-th powers of the roots of the (conjugate of the) characteristic polynomial $c X^{a+1}-(1+c) X^a + 1$, where $c=2 q/p \in (0,\infty)$. This polynomial is of degree $a+1$ and has $1$ as a root. This last property is also true for the characteristic polynomials of the relations for the $\pi_k$ for $k$ in $\ll i_s+2,i_s+a\rr$ or $> i_s + a + 1$. As the quartic equation has explicit solutions, the expressions of the $\pi_k$, $k\ge a$ are fully explicit if $a+1 \le 4 + 1$, that is $a\le 4$. In particular for $a=1$, the $(\pi_k)_{k\in \ll 1, i_s \rr}$ and $(\pi_k)_{k> i_s }$ are two geometric progressions:  

\begin{equation}\label{pik1}
\forall k \in \llbracket 1, i_s \rrbracket, \, \pi_k = c ^k \pi_0, \qquad
\forall k>i_s, \,  \pi_k = p(1-p)^k \left(\frac{2}{p}\right)^{i_s+1}  \pi_0.
\end{equation}

For $a > 4$ the roots of these polynomials can be found numerically.

\section{Choice of $i_s$} \label{app:choiceis}

\begin{figure}[h!]
\centering
\includegraphics[scale=1]{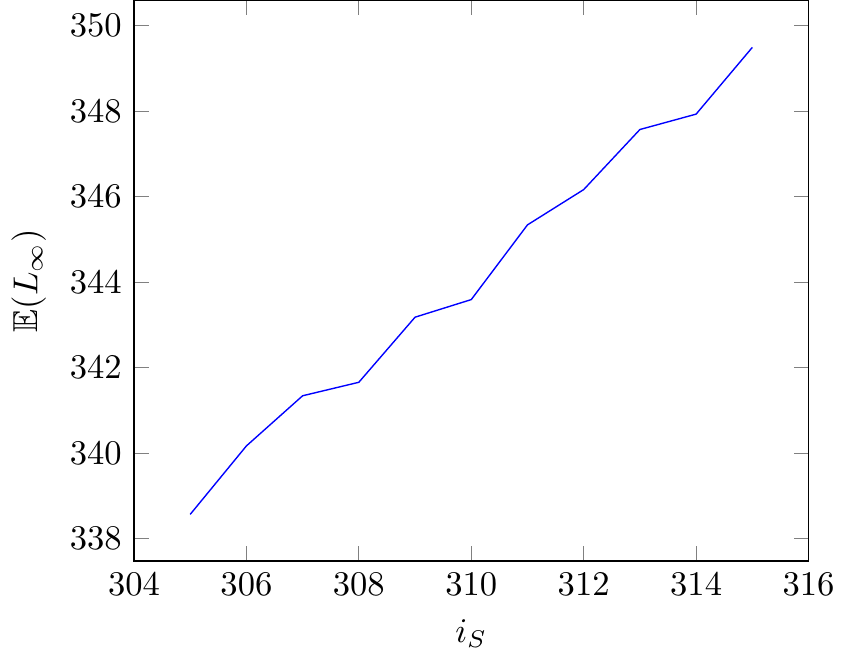}
\caption{\footnotesize{The expected length $\E(\Linf)$ as function of $i_s$.}}
\label{fig:estimateIs}
\end{figure}

To rigorously compare the variance of the simplified model~\eqref{eq:model} with the one of~\eqref{eq:modelcomplet}, we choose $i_s$ so that the ceiling function of the mean of the equilibrium of~\eqref{eq:model} $\lceil\E(\Linf)\rceil$  is the same as the one for~\eqref{eq:modelcomplet}, \emph{i.e.} $342$ bp in~\cite{Xu13}. We take the biological parameters obtained in~\cite{T04,Soudet} and used in~\cite{Xu13}:
 \[a = 7,\, p = 0.026,\,  L_s=90, \, \beta = 0.045,\]
where $\beta$ is a fitting parameter describing telomerase recruitment to telomeres~\cite{Xu13} (Table \ref{table-param}).
In order to compute the corresponding mean of the equilibrium, for each choice of $i_s$ we run $10^6$ numerical simulations of~\eqref{eq:model}. This mean is then computed and plotted as a function of $i_s$, \emph{cf.} Fig. \ref{fig:estimateIs}. Finally, we chose the value of $i_s$ that gives $\lceil\E(\Linf)\rceil = 342$ bp. This procedure leads to $i_s = 308$ bp.

\section{Expected Time of Senescence} \label{app:expected}

The aim of this section is to establish formula \eqref{eq:PTgen}. 
Before entering the details just note that taking $a=1$ is not strictly equivalent to make a change in the length unit as the lenghts of the telomeres are integer-valued. But taking $a>1$ only make the results more difficult to state without exhibiting a new behaviour.
For the sake of simplicity, we drop the superscript $i$ in this section and start by studying a typical couple $(X_n, Y_n)$. The shortening of these two telomeres can be mathematically translated into the following model:
\begin{equation}\label{eq:coupledModel}
\begin{pmatrix}
X_{n+1}\\
Y_{n+1}
\end{pmatrix}
=
\begin{pmatrix}
(X_{n}- a\cdot \B_n)^+\\
(Y_{n} - a\cdot (1-\B_n))^+
\end{pmatrix},
\end{equation}
where $B_n$ is a Bernoulli random variable of parameter $1/2$, and $(X_0,Y_0) \overset{dist.}{\sim} \Pi$.

This process is an oriented simple random walk on $\mathbb{Z}^2$ until one of the coordinates reaches zero, and can be written explicitly:
\begin{equation} \label{eq:XnYnexpl}
\begin{aligned}
  X_n &= X_0 - a \sum_{k=1}^{n} \B_k = X_0 - a \Bin (n,1/2), \\
  Y_n &= Y_0 - a \sum_{k=1}^{n} (1-\B_k) = Y_0 - a \left( n - \Bin (n,1/2) \right), 
\end{aligned}
\end{equation}
where $\Bin(n,1/2)$ is a binomial distribution of parameters $n$ and $1/2$.
In this case, let us define the first time one of the coordinates reaches zero, $T^1$, as:
\[T^1 = \inf \left\{ n \ge 0, \min(X_n,Y_n) < 0 \right\}\]
Then, from \eqref{eq:XnYnexpl} and for $a=1$, because $(X_n)$ and $(Y_n)$ are non-increasing, we get:
\begin{equation} \label{calcul}
\begin{split}
\mathbb{P}(T^1 > n) &= \P ( X_n \ge 0,Y_n \ge 0) = \P (n  - Y_0  \le \Bin(n,1/2)  \le X_0  ) \\
&= \sum_{\substack{k+l \ge n \\k,l\ge0 }}\Pi(X_0 = k,Y_0 = l) \; 2^{-n} \sum_{t= n - l }^{k} \binom{n}{t}.
\end{split}
\end{equation}
From here, we easily derive the distribution of the time of senescence by considering all $16$ independent pairs of telomeres:
\begin{equation} \label{calcul2}
\begin{split}
\P(T > n)&=\P( \neg\text{\{senescence at the $n^{th}$ generation\}} )\\
&= \P\left( \min_{1\le k \le 32} L^n_k \ge 0\right) = \P(\forall i \in \llbracket 1,16 \rrbracket, \min (X_n^i, Y_n^i) \ge 0 )\\
&= \P(\min (X_n, Y_n) \ge 0 )^{16} = \mathbb{P}(T^1 > n)^{16}.
\end{split}
\end{equation}

Formulas \eqref{calcul} and \eqref{calcul2} lead to \eqref{eq:PTgen}, which gives the expected time of senescence \eqref{eq:expT} using $\E(T) = \sum_{n=0}^{\infty} \P(T>n)$.

\section{Asymptotics of the Mean Time of Senescence} \label{app:asymp}

The aim of this section is to prove formulas \eqref{eq:loideT1} and \eqref{eq:expasymp}. To do so we first note that, using a generalized version of the central limit theorem, for any fixed positive number $t$, and any sequence $u_n \sim tn$:
\begin{equation} \label{Lucas}
\frac{\Bin(u_n,\frac12) - \frac12 u_n}{\sqrt{n}}
\overset{\mathcal{L}}{\underset{n\rightarrow +\infty}{\longrightarrow}}
\frac12 \, \Normal(0,t),
\end{equation}
where $\Bin(u_n,\frac12)$ is a binomial distribution with parameters $u_n$ and $1/2$.






Let $w$ be a real number. As in \eqref{calcul}, we get, with $n(x_0)= \left\lfloor 2x_0 - w \sqrt{x_0} \right\rfloor \sim 2 x_0 $:
\begin{equation*} \label{calcul3}
\begin{split}
& \P \left(\frac{2 x_0 - T^1_{x_0}}{\sqrt{x_0}} < w \right)=\P \left(T^1_{x_0} > 2x_0 - w \sqrt{x_0} \right) = \P \left(T^1_{x_0} > n(x_0) \right)\\ 
&= \P \left( n(x_0) - x_0 \le \Bin\left(n(x_0),\frac12 \right) \le x_0 \right)
= \P \left( \abs{\frac{\Bin\left(n(x_0),\frac12 \right) - \frac{n(x_0)}2}{\sqrt{x_0}}} \le \frac{x_0 - \frac{n(x_0)}2}{\sqrt{x_0}} \right).
\end{split}
\end{equation*}


As $x_0$ tends to $\infty$, thanks to \eqref{Lucas}, we obtain:
\[
\P \left(\frac{2 x_0 - T^1_{x_0}}{\sqrt{x_0}} \le w \right) \to \P\left(\frac{\abs{N}}{2} \le \frac{w}{2} \right),
\]
where $N \sim \Normal(0,2)$.

\medskip

The speed of convergence is given by the Berry-Essen theorem in terms of the cumulative distribution function of $\frac{2 x_0 - T^1_{x_0}}{\sqrt{x_0}}$, say $F_{x_0}$:
\[
\forall x\in \R, \forall x_0 \in \N \qquad \abs{F_{x_0}(x) - \erf x} \le C \, x_0^{-1/2},
\]
where $C$ is an absolute constant. This inequality should provide a bound to the difference in equation~\eqref{eq:expasymp}:
\[
\abs{\E(T_{x_0}) - \left(x_0 +\sum_{k=0}^{x_0-1}  \left[  \erf \left( \frac{k}{2 \sqrt{x_0}}\right) \right]^{16}\right)}.
\]

\newpage

\section{Parameters used in this study} \label{app:table}

\begin{table}[h!]
\begin{center}
\begin{tabular}{|c|m{55mm}|}
\hline
$a$ & Length of the 3'-end overhang. \\ \hline
$S$ & Threshold length of the shortest telomere inducing senescence. \\ \hline
$L_s$ & Length threshold for the function $f$, below which telomerase is recruited to the telomere with probability $1$.\\ \hline
$i_s$ & Length threshold for the simplified model of telomerase recruitment. \\ \hline
$p$ & Parameter of the geometric random variable $\G_n$. \\ \hline
$\beta$ & Parameter of the function $f$, fitted on experimental data~\cite{Xu13}. \\
\hline
\end{tabular}
\end{center}
\caption{\footnotesize{Parameters used in this study.}}
\label{table-param} 
\end{table}

\end{appendices}

\vfill

\medskip
\noindent{\bf Acknowledgments.} The authors thank Marie Doumic, Lucas Gerin, Philippe Robert and Maria Teresa Teixeira for their critical reading of the manuscript. Z. X.'s research is supported by the Initiative d'Excellence program (Grant DYNAMO, ANR-11-LABX-0011-01), the Edmond de Rothschild foundation and Fondation pour la recherche m\'edicale (FRM équipe labellisée DEQ20160334914 to the Teixeira lab). S. E.'s research is supported by ERC Starting Grant SKIPPER$^{AD}$ No. 306321. T. B.'s research is supported by the European Research Council (ERC) under the European Unions Horizon 2020 research and innovation programme (grant agreement No 639638).

\end{document}